# MESOPAUSE TEMPERATURE RETRIEVAL FROM SATI MEASURED SPECTRA WITH AN LINEAR BACKGROUND COMPONENT. RESULTS FROM THE NUMERICAL SIMULATION.

*Atanas Marinov Atanassov*

*Space and Solar-Terrestrial Research Institute, department in Stara Zagora;*
*E-mail: At_M_Atanassov@yahoo.com*

**Abstract**
   The Spectral Airglow Temperature Imager is a ground-based spectral instrument for spatial registration of airglow emissions. The basic aim of the instrument development is the investigation of gravity waves based on the spatial characteristics of the temperature field at the altitude of mesopause and its evolution in the time. The temperature retrieval is based on matching measured and preliminary calculated synthetic spectra.
   Possibilities are presented for generalization of the basic regression equation which connects the measured and the synthetic spectra. A linear change of the background for the entire filter transmittance interval was presumed. Numerical experiments by Monte-Karlo simulation were conducted.
   The presented results show a bigger stability of the proposed approach in comparison with the traditional one, without considering the linear background.

   **Introduction**
   The Spectral Airglow Temperature Imager (SATI) is a ground-based Fabry-Perot spectrometer for spatial registration of airglow emissions [1, 2]. The SATI instrument was originally intended for investigation of internal gravity waves, propagating at the altitude of the mesopause [3]. This is possible by processing the registered images (Fig.1a). Twelve sector spectra are determined from each image (fig. 1b, c). These spectra are compared with synthetic spectra, calculated in advance for different temperatures for determination of the mesopause temperature where the respective emissions are radiated. The radiation maximum of these airglow emissions is disposed at a mean altitude of 97 km for the transitions $O_2(b^1\Sigma_g^+ - X^3\Sigma_g^-)$. The processing of images from one night séance yield the rotational temperature and the emission rate sector time series [1, 2] determined for different points from the annular sky segment at the altitude of the mesopause. The propagation and the cinematic parameters of the internal gravity waves are determined on the basis of the analysis of these time series [3].

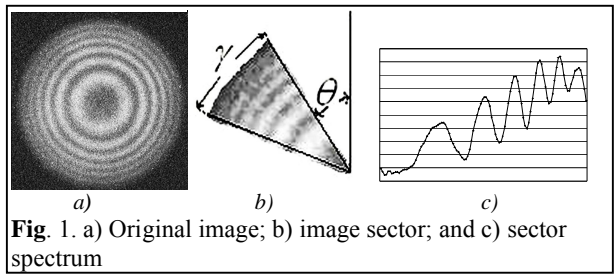
*a)    b)    c)*
**Fig**. 1. a) Original image; b) image sector; and c) sector spectrum

   A version of SATI- 3SZ was manufactured at the Department of the Solar-Terrestrial Influences Laboratory in Stara Zagora [4] in collaboration with the Space Instrumentation Laboratory at the Centre for Research in Earth and Space Science (CRESS), York University, Toronto, Canada. Additionally, algorithms for retrieval of the rotational temperature from the registered $O_2$ spectra were investigated and developed [5-6].

**Temperature retrieval**

The determination of the mesopause temperature by SATI-registered images is based on the comparison of the measured spectra with a set of patterns – calculated in advance for different values of the temperature of the radiated oxygen molecules spectra which correspond to transitions $O_2(b^1\Sigma_g^+ - X^3\Sigma_g^-)$. These transitions correspond to wave longitude in the interval 864÷868 nm. The maximum radiation of the SATI-registered signal is at the height of the mesopause. The comparison of the measured spectrum M with the synthetic spectra $S_T$, calculated in advance for different temperatures for long-time night interval () allows to determine the equilibrium temperature of the oxygen molecules, radiating the signal, incoming toward the Earth. This is done on the basis of the following linear regression equation:

(1) $\quad M = E.\widetilde{S}_{T_{rot}} + B$

In the last equation, the physical meaning of the multiplicative coefficient E is an integral emission intensity for all filter transmittance intervals. The meaning of B is a mean intensity of the background. The solution of regression equation (1) yields the values of E and B. The rotational temperature $T_{rot}$ is determined on the basis of minimizing the functional

(2) $\quad \delta_{L,T_{rot}} = \sqrt{\dfrac{1}{p_2 - p_1} \sum_{p=p_1}^{p_2}(M_p - E.\widetilde{S}_{T_{rot},p} - B)^2 . w_p}$

The additive constant B in regression equation (1) allows the admission of even background intensities for the entire spectral interval. The weight coefficients $w_p$ satisfy condition $\sum_p w_p = 1$.

**Form of the basic regression equation in the case of a background with linear changing intensity**

The conducted measurements are accepted to be correct and usable when a low background signal is retrieved. Besides, the background signal is accepted even for the entire filter transmittance spectral interval. In all cases when spectral pollution from a secondary source is available, the respective measurements are eventually removed.

We will accept a linear change of the background by frequency and generalization of the basic regression equation (1)

(3) $\quad M = E.\widetilde{S}_{T_{rot}} + B^0 + B^{\angle}$

The background B was presented by two components $\overline{B}$ and $B^{\angle}$ in equation (3). Component $\overline{B}$ is even by intensity for the entire spectral interval of the background which was passed from the filter; component $B^{\angle}$ is presumed dependant linearly in the frame of the same spectral interval. Both components can be changed for the time of every next measurement as a result of the change of the direction of the incoming signal in relation to the stars or some kinds of artificial sources. Component $B^{\angle}$ is unknown in the space and time and can be determined by minimization of the generalized version of (2)

(4) $\quad \delta_{L;T_{rot}} = \sqrt{\dfrac{1}{p_2 - p_1} \sum_{p=p_1}^{p_2}(M_p - E.\widetilde{S}_{T_{rot},p} - B^0 - B^{\angle})^2 . w_p}$

The solution of regression task (3) by the same tools as (1) is possible after transformation to

(5) $\quad M - B^{\angle} = E.\widetilde{S}_{T_{rot}} + B^0$

or

(5') $\quad M^{\angle} = E.\widetilde{S}_{T_{rot}} + B^0$,

In (5') $M^{\angle}$ is a modified measured spectrum. The minimization is based already on

(6) $\quad \delta_{L;T_{rot},B^{\angle}} = \sqrt{\dfrac{1}{p_2 - p_1} \sum_{p=p_1}^{p_2} (M_p^{\angle} - E.\widetilde{S}_{T_{rot},p} - B^0)^2 . w_p}$

In the new regression task, the rotational temperature is found when (6) has a minimum along with the additional condition for background with linear intensity.

The reasons for admission of uneven background are purely physical and are connected with the natural conditions of the measurement. Except the natural airglow emissions, other sources of a natural background are also available.

**Monte-Karlo simulation**

A series of simulations was performed as for each temperature in the interval (110÷300°K) one "measured" spectrum was generated. This was made as the relative synthetic spectrum $S_{T_{rot}}$ was transformed towards the image space by choosing typical values of the filter parameters (refractive index and central wavelength) in order to conduct the entire experiment.

(7) $\quad \widetilde{S}_{T_{rot},p} = \sum_{\lambda_p}^{\lambda_p + 1} S_{T_{rot},\lambda}$

In the last formula $\lambda_p$ is the wavelength, corresponding to the p-th pixel in the one-dimensional space of the measured spectra [5]. A multiplicative gaining coefficient E is applied; with this coefficient the values of the spectrum for every pixel are fitted to the real ones:

(8) $\quad M_{T_{rot},p} = E.\widetilde{S}_{T_{rot},p}$

A random noise was added to the spectrum with a generator of random numbers $\phi(p) = \Phi.(RANDOM - .5)$ with amplitude $\Phi$. Two background components - even and uneven, are added to the spectrum, thus obtained, and we finally have

(9) $\quad M_{T_{rot},p} = E.\widetilde{S}_{T_{rot},p} + \overline{B} + B^{\angle} + \phi_p$

The aim of the investigation is to establish the behavior of the original approach for temperature retrieval without taking into account component $B^{\angle}$ as well as the behavior of the modified one, taking into account this component.

For each temperature spectrum, generated in the observed interval (110-300K), "polluted" with uneven background, 20 tests were performed for temperature retrieval by the two approaches.

Uneven background with linear change and different intensity at both ends b2= b1 + b(rand-.5) was generated by every test. Value b1 is the background intensity at the first end of the spectrum, however, at the other end we have b2= b1

**Table 1.**

| E= 25 | E= 50 | E= 100 |
|---|---|---|
| Noise= 5 | Noise= 5 | Noise= 5 |
| B=(20;20) | B=(20;20) | B=(20;20) |
| E= 25 | E= 50 | E= 100 |
| Noise= 5 | Noise= 5 | Noise= 5 |
| B=(20;15) | B=(20;15) | B=(20;15) |
| E= 25 | E= 50 | E= 100 |
| Noise= 5 | Noise= 5 | Noise= 5 |
| B=(20;5) | B=(20;5) | B=(20;5) |

+ b.(rand-.5); where b is the doubled maximum of the possible difference of the background amplitude at this end and $\xi$- is the random number within the interval (-.5,5). A series of experiments was conducted with different values of even and uneven backgrounds ($\overline{B}$, $B^{\neq}$)=(20,20; 20,15; 20,5). Experiments with different values of the random noise amplitude $\Phi$ =(1;3;5) and different gain coefficients E=(25,50,100) were conducted.

### Results from the Monte-Karlo simulation experiments

Figures 2 and 3 contain histograms which present the temperature retrieval process by applying the two models M and B. The initial temperatures, at which "the measured spectra" were generated, were 170°K for fig.2 and 210°K for fig.3 by intensities 25, 50 and 100 respectively. During these experiments, 1000 tests were performed for temperature retrieval with addition of random noise and background with linear intensity. The difference between the two approaches is obvious. The retrieval temperatures were distributed in the ranges 67° and 95°, respectively, for initial temperatures of 170° and 210° and with model intensity 25. At higher intensity of the modeled signal of 50 units, the intervals are narrowed to 33°K and 46°K, respectively, and to 16°K and 22°K at intensity of 100, which correspond to reduction of the significance of the background with linear intensity. When applying the temperature retrieval model B, the intervals in which the temperatures are distributed at intensity 25, are 28°K and 57°K, 14°K and 19°K at intensity 50 and 7°K and 10°K at intensity 100, respectively.

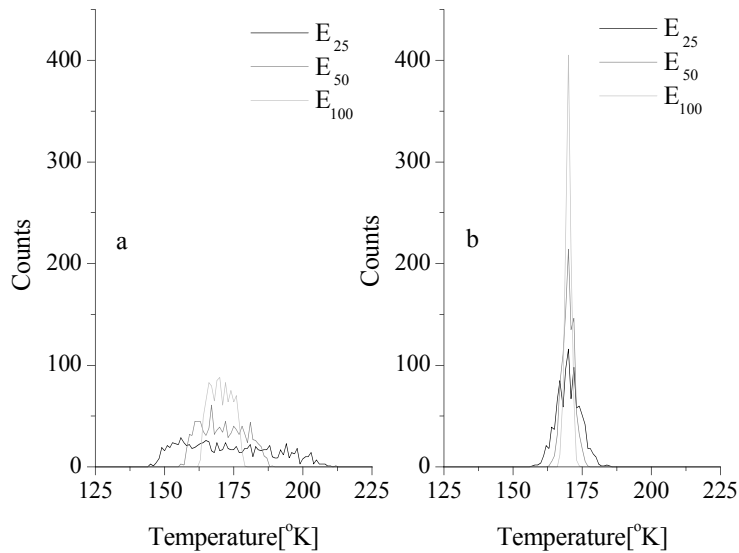

**Figure 2.** Histograms of retrieved temperatures by original temperature 170°K-
(a) by original method and (b) by uneven background reduction

Figures 4 and 5 show the mean square deviation and the absolute errors for the retrieved values of the temperature in series from 20 tests for each initial temperature in the interval (110-300°K). Figures a-c refer to intensity E of 25 and to values of the background component with linear intensity of 20, 15 and 5, respectively. The intensity is 50 units in the figures from the second column (d-f) and in the third one

(g-i) - 100 units, with analogous background. It is obvious that the retrieved values fall within a very large interval, without considering the background with linear

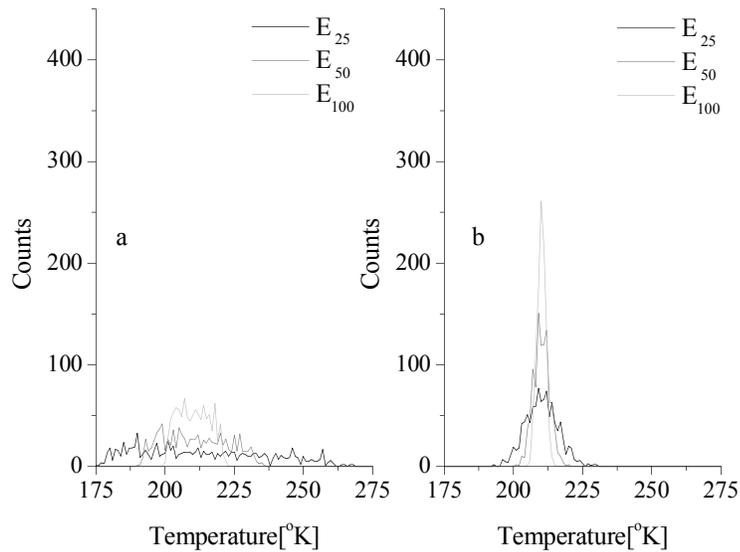

Figure 3. Histograms of retrieved temperatures by original temperature 210°K- (a) by even component only and (b) by linear background

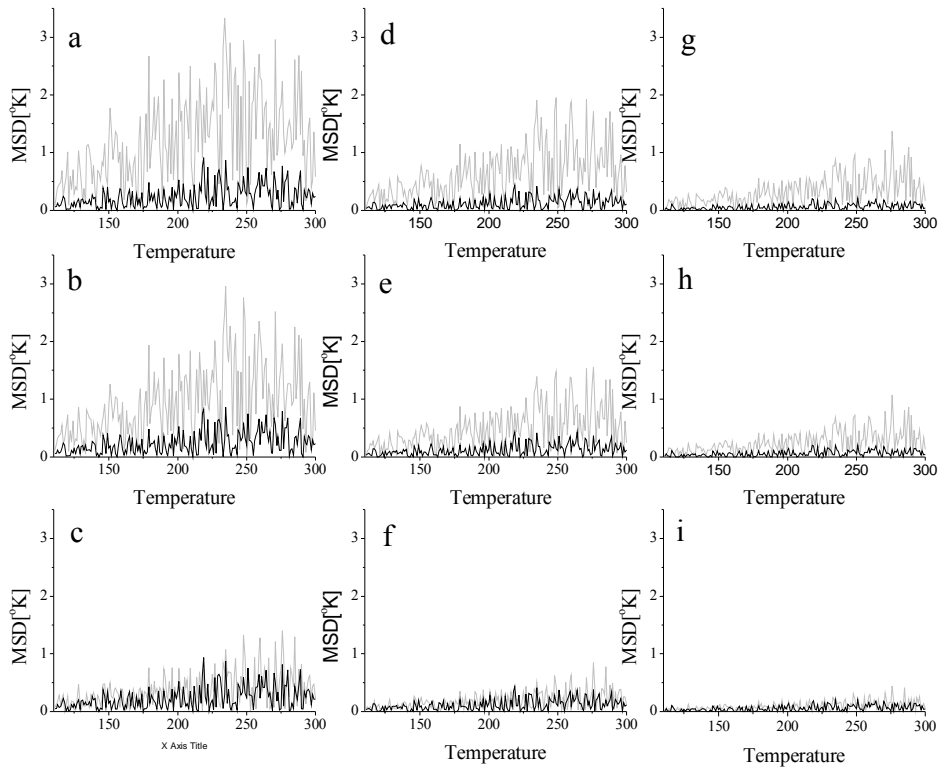

Figure 4. Mean Standard deviation for E=25 and B< (20,15,5) respectively (a-c); analogously E= 50 and B< (20,15,5) respectively (d-f) and E= 100 and B< (20, 15, 5) (g-i).

intensity. The values which were retrieved by model B fall within a rather narrow interval.

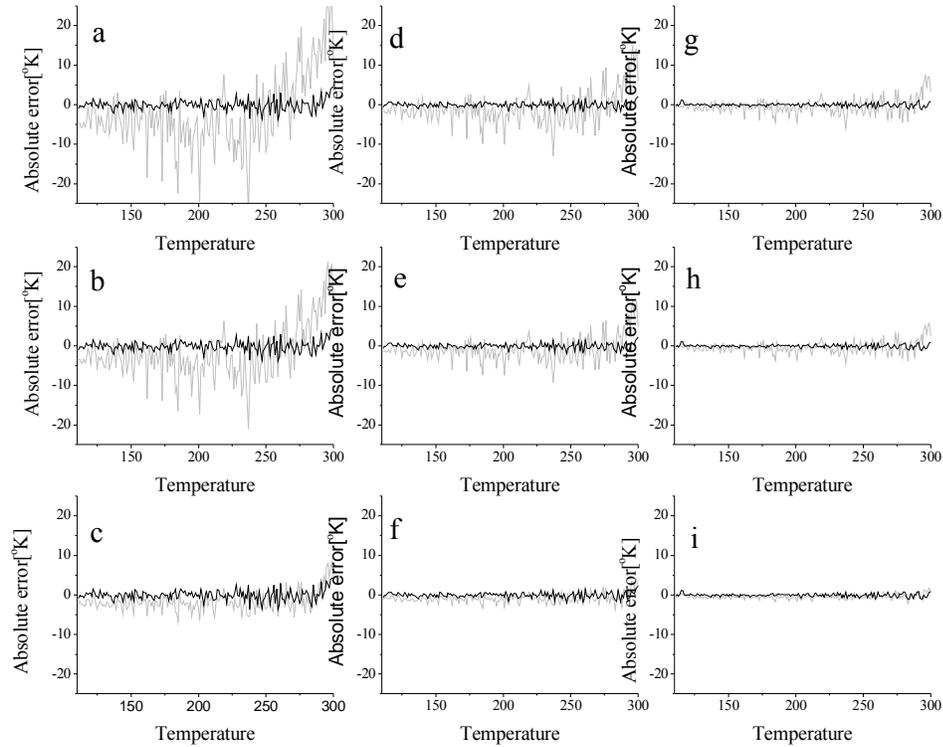

Figure 5. Absolute error for E=25 and B< (20,15,5) respectively (a-c); analogously E= 50 and B< (20,15,5) respectively (d-f) and E= 100 and B< (20, 15, 5) (g-i).

The absolute error grows by applying of model M with increasing of the retrieved temperature. Respective increasing by applying of model B is minimal.

Some asymmetry in absolute errors for model M is obviously, which speak probably for potential systematic error by retrieving of temperatures

**Conclusions**

If a background component with linear intensity is available in the measured spectra, the proposed approach would yield better temperature results. In all cases, the mean results from the temperature retrieval, taking into consideration the uneven background, will be in the range of one degree for almost the entire temperature interval of the simulation (110-300K).

Except the development of the basic calculation model with addition of uneven background, which is random in the time by its character, very interesting appears a similar development with a significant growth of the background, connected with additional plank sources. The investigation of the spectral pollution with a solar spectrum during sunrise or sunset, or the Moon reflected solar light when the Moon is over the horizon, are current tasks, focusing the attention.

The application of the present approach for minimization of errors by real data is a very attractive idea, aimed at decreasing the noise in the nocturnal course of the sector temperatures, determined by SATI-registered images. In real conditions, along with the possible linear background, most probably other parameters of the calculation models will also have a random character. Thus, the errors of the filter parameters determination will lead to errors of the synthetic spectra transformation [5]. These errors represent specific discrepancies between the transformed synthetic and measured spectra. Therefore, the errors, connected with the background component with linear intensity, will compete with the noise in the measured spectrum and the unsystematic random noise of the filter parameters determination. The analysis of the influence of these errors is a possible next stage of the investigation.

**Acknowledgment**